\documentclass[iop]{emulateapj}

\newcommand\science{{Science}}%
\newcommand\ijmpe{{International Journal of Modern Physics E}}%

\usepackage[bookmarks,colorlinks,citecolor=cyan]{hyperref}
\usepackage{ulem}
\usepackage{graphicx}

\def\ngc#1{\hbox{NGC\,#1}}
\def\feh{\hbox{\rm [Fe/H]}}
\def\hst{{\it HST\/}}

\def\afe{\hbox{\rm [$\alpha$/Fe]}}
\newcommand{\msun}{$M_{\odot}\,$}
\slugcomment{}

\shorttitle{Kinematic separation for the Bulge globular \ngc6528}
\shortauthors{E.~P. Lagioia et al.}

\begin{document}

\title{On the kinematic separation of field and cluster stars across the Bulge
globular \ngc6528}

\author{
 E.~P.~Lagioia\altaffilmark{1},
 A.~P.~Milone\altaffilmark{2,3,4},
 P.~B.~Stetson\altaffilmark{5},
 G.~Bono\altaffilmark{1,6},
 P.~G.~Prada Moroni\altaffilmark{7}
 M.~Dall'Ora\altaffilmark{8},
 A.~Aparicio\altaffilmark{3,4},
 R.~Buonanno\altaffilmark{1,9},
 A.~Calamida\altaffilmark{10,6},
 I.~Ferraro\altaffilmark{6},
 R.~Gilmozzi\altaffilmark{11},
 G.~Iannicola\altaffilmark{6},
 N.~Matsunaga\altaffilmark{12},
 M.~Monelli\altaffilmark{3,4},
 A.~Walker\altaffilmark{13}
 }

\altaffiltext{1}{Dipartimento di fisica, Universit\`a degli studi di Roma -- Tor
Vergata, via della Ricerca Scientifica 1, I-00133, Roma, Italy;
email:eplagioia@roma2.infn.it}
\altaffiltext{2}{Research School of Astronomy and Astrophysics, The Australian
National University, Cotter Road, Weston, ACT, 2611, Australia}
\altaffiltext{3}{Instituto de Astrof\`isica de Canarias, E-38200 La Laguna,
Tenerife, Canary Islands, Spain}
\altaffiltext{4}{Department of Astrophysics, University of La Laguna. E-38200 La
Laguna,
Tenerife, Canary Islands, Spain}
\altaffiltext{5}{Dominion Astrophysical Observatory, Herzberg Institute of
Astrophysics, National Research Council, 5071 West Saanich Road, Victoria, BC
V9E 2E7, Canada}
\altaffiltext{6}{INAF - Osservatorio Astronomico di Roma, Via Frascati 33,
I-00044 Monte Porzio Catone, Italy}
\altaffiltext{7}{Dipartimento di Fisica, Universit\`a di Pisa, I-56127 Pisa,
Italy}
\altaffiltext{8}{INAF - Osservatorio Astronomico di Capodimonte, Salita
Moiariello 16, I-80131 Napoli, Italy}
\altaffiltext{9}{INAF - Osservatorio Astronomico Collurania, via M. Maggini,
I-64100 Teramo, Italy}
\altaffiltext{10}{Space Telescope Science Institute, 3700 San Martin Dr.,
Baltimore, MD 21218, USA}
\altaffiltext{11}{European Southern Observatory, Karl-Schwarzschild-Stra{\ss}e
2,
85748, Garching, Germany}
\altaffiltext{12}{Kiso Observatory, Institute of Astronomy, School of Science,
The University of Tokyo, 10762-30, Mitake, Kiso-machi, Kiso-gun,3 Nagano
97-0101, Japan}
\altaffiltext{13}{Cerro Tololo Inter-American Observatory, National Optical
Astronomy Observatory, Casilla 603, La Serena, Chile}

\begin{abstract}
We present deep and precise multi-band photometry of the Galactic Bulge globular
cluster \ngc6528. The current dataset includes optical and near-infrared images
collected with ACS/WFC, WFC3/UVIS, and WFC3/IR on board the Hubble Space
Telescope.  The images cover a time interval of almost ten years and we have
been able to carry out a proper-motion separation between cluster and field
stars. 

We performed a detailed comparison in the $m_{\rm F814W}, m_{\rm F606W} - m_{\rm
F814W}$ Color-Magnitude Diagram with two empirical calibrators observed in the
same bands. We found that \ngc6528 is coeval with and more metal-rich than
47\,Tuc. Moreover, it appears older and more metal-poor than the
super-metal-rich open cluster \ngc6791. The current evidence is supported by
several diagnostics (red horizontal branch, red giant branch bump, shape of the
sub-giant branch, slope of the main sequence) that are minimally affected by
uncertainties in reddening and distance.  

We fit the optical observations with theoretical isochrones based on a
scaled-solar chemical mixture and found an age of $11\pm1$\,Gyr and an iron
abundance slightly above solar ($\feh = +0.20$).  The iron abundance and the old
cluster age further support the recent spectroscopic findings suggesting a rapid
chemical enrichment of the Galactic Bulge.
\end{abstract}


\section{Introduction}\label{sec:int}
In one of the three seminal papers given by~\citet{Baa58} at the famous Vatican
conference held in Rome in 1957, he discussed the stellar populations in the
Galactic Bulge. It is interesting to note that Baade called the same region van
Tulder's pole. \citet{vant}, a few years before, published an interesting paper
concerning the location of the Galactic pole by using different types of stars
and open clusters. Thus, we suggest that this low-reddening region should be
defined as the van Tulder-Baade's (vTB) window.

This region is dominated by the presence of two Galactic globular clusters
(GGCs), namely \ngc6528 and \ngc6522. The former is a very interesting cluster,
since it is among the most metal-rich GGCs. Spectroscopic investigations, based
on high-resolution spectra, suggest for this cluster a solar metallicity and a
modest $\alpha$-element enhancement. \citet{Car01}, studying four red Horizontal
Branch (HB) stars, found $\feh = +0.07 \pm 0.01$ and a marginal $\alpha$
enhancement ($\afe \approx +0.2$), while \citet{Zoc04} investigated three giants
belonging to HB and to the Red Giant Branch (RGB) and found $\feh = -0.1 \pm
0.2$ and $\afe \approx +0.1$. In a similar investigation, based on
high-resolution near-infrared (NIR) spectra of four bright giants, \citet{Ori05}
found $\feh = -0.17 \pm 0.01$ and $\afe = +0.33 \pm 0.01$. The above measurements
indicate that \ngc6528 is an ideal laboratory to constrain the possible
occurrence of an age-metallicity relation among the most metal-rich and old GGCs
\citep{Rak05,Dot11}. Moreover, \ngc6528 and its twin \ngc6553 are considered the
prototype metal-rich clusters of the Galactic Bulge \citep{Ort95,Zoc01} and
therefore, unique stellar systems to constrain the difference between cluster
and field stars.

Estimates of both structural parameters and intrinsic properties for \ngc6528
are partially plagued by the high differential reddening across the field of
view (FoV) and by the strong level of contamination from Bulge and Disk stars. 

The effect of differential reddening and field-star contamination are among the
reasons why age estimates for \ngc6528 range from $13 \pm 2$\,Gyr \citep[][based
on isochrones provided by \citeauthor{Cas97} \citeyear{Cas97} and
\citeauthor{Cast99} \citeyear{Cast99}]{Ort01} to $11\pm2$\,Gyr
\citep[][isochrones by \citeauthor{Sal00} \citeyear{Sal00}]{Fel02}, to 12.6\,Gyr
\citep[][isochrones by \citeauthor{Ber94} \citeyear{Ber94}]{Mom03}.

In principle, the combination of visual and NIR photometry is an efficient
diagnostic to separate candidate cluster and field stars \citep{Cal09,Bon10}.
Indeed, in many GGCs, field stars and cluster members use to populate different
regions of the optical-NIR color-color plane.  Unfortunately, the metallicity
distribution of \ngc6528, Bulge and thin-disk stars all peak around solar
chemical composition, thus making \ngc6528 almost indistinguishable from the
field in many color-color plane.

An alternative approach to separate field and cluster members is based on the
analysis of stellar proper motions.  A partial separation of \ngc6528 field and
cluster stars was performed for the first time by \citet{Fel02} by using
accurate astrometric measurements from  multi-epoch images taken with the
Wide-Field Planetary Camera 2 (WFPC2) on board the {\it Hubble Space Telescope}
(\hst).  A similar approach was also adopted by \citet{Zoc01} to constrain
the Bulge membership of \ngc6553.

In this investigation, we exploit the unique astrometric and photometric
performance of the Ultraviolet and Visual Channel of the Wide Field Camera 3
(UVIS/WFC3), the Infrared Channel of WFC3, (IR/WFC3) and of the Advanced Camera
for Survey (ACS) of {\it HST} to obtain high-accuracy astrometry and photometry
of stars in the \ngc6528 FoV. Proper motions derived from this dataset
allowed us to separate, with unprecedented precision, most of the field stars
from cluster members. The field-stars decontamination of the CMD allowed us to
correct the photometry of \ngc6528 for differential reddening, hence to derive
an accurate age estimate by comparing the CMD of \ngc6528 with isochrones. 

The structure of the paper is the following. In \S 2 we present both the optical
and the NIR data sets used in the current investigation together with the
approach adopted to perform the photometry. The approach adopted to measure the
proper motion and to select candidate cluster stars are discussed in subsection
2.1, while the subsection 2.2 deals with the method adopted to estimate the
differential reddening. 

The selection of the candidate field stars is presented in \S 3, while in \S 4
we discuss the optical CMDs of candidate field and cluster stars together with
advanced evolutionary features (red horizontal branch, RGB bump). In \S 5 we
discuss in detail the comparison with two empirical calibrators: the old,
metal-rich globular 47\,Tuc and the super-metal-rich open cluster \ngc6791. \S 6
deals with the comparison with cluster isochrones and the difference with the
age and the metallicity of the empirical calibrators. A brief summary of the
results and a few possible avenues of the current project are outlined in \S 7. 

\begin{figure*} 
\centering
\includegraphics[width=10cm]{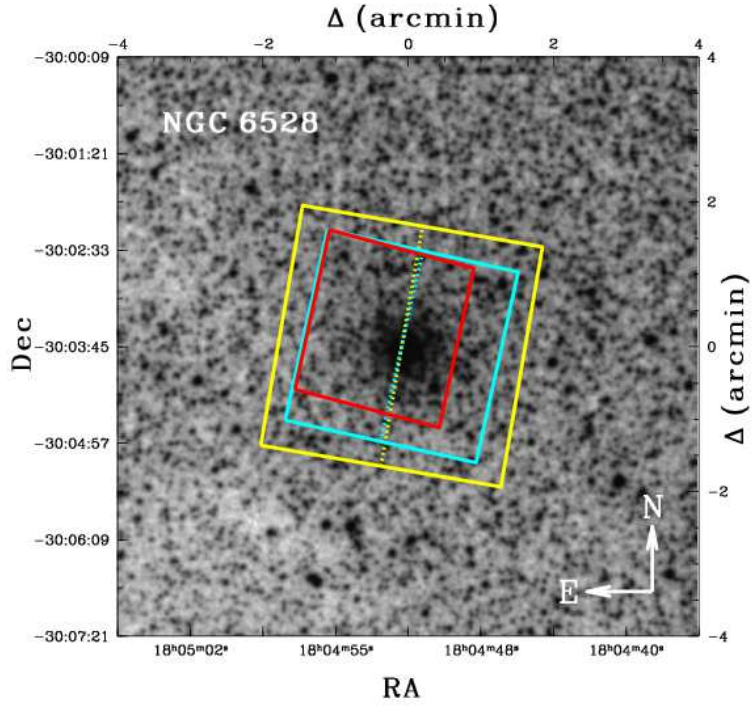}
\caption{Footprints of optical and NIR \hst\ images superimposed on a
8$\arcmin\times\,$8$\arcmin$ DSS red POSS-II plate.  The yellow, cyan and red boxes
display the ACS/WFC, the WFC3/UVIS and the WFC3/IR pointings. The orientation is
North up and East left.} 
\label{fig:fp}
\end{figure*}


\section{Observations and data reduction}\label{sec:obs} 
The data presented in this investigation come from \hst\ archive images centered
on \ngc6528. The Advanced Camera for Survey / Wide Field Camera (ACS/WFC)
dataset was collected on June 6th 2002~\footnote{Program GO\,9453,
PI.\,T.\,Brown.}, in the filters $F606W$ ($1\times4\,{\rm s}, 1\times50\,{\rm
s}, 1\times450\,{\rm s}$) and $F814W$ ($1\times4\,{\rm s}, 1\times50\,{\rm s},
1\times450\,{\rm s}$). The Wide Field Camera 3 (WFC3) dataset was collected
between June 26th and 27th 2010 and includes optical and NIR images. The former
were taken with the Ultraviolet and Visual channel (WFC3/UVIS)~\footnote{Program
GO\,11664, PI.\,T.\,Brown.} in the filters $F390W$ ($2\times40\,{\rm s},
2\times348\,{\rm s}, 2\times715\,{\rm s}$), $F555W$ ($1\times1\,{\rm s},
1\times50\,{\rm s}, 1\times665\,{\rm s}$) and $F814W$ ($1\times1\,{\rm s},
1\times50\,{\rm s}, 2\times370\,{\rm s}$). The latter were taken with the WFC3 -
Infrared channel (WFC3/IR) in the filters $F110W$ ($3\times49\,{\rm s},
1\times299\,{\rm s}, 2\times399\,{\rm s}$) and $F160W$ ($3\times49\,{\rm s},
1\times299\,{\rm s}, 2\times399\,{\rm s}$).  The footprints of the different
datasets are shown in Fig.~\ref{fig:fp}.

In addition, to compare the CMD of \ngc6528 with that of the open cluster
\ngc6791  we used ACS/WFC data in $F606W$ ($1\times0.5\,{\rm s}, 1\times5\,{\rm
s}, 1\times50\,{\rm s}$) and $F814W$ ($1\times0.5\,{\rm s}, 1\times5\,{\rm s},
1\times50\,{\rm s}$) for \ngc6791~\footnote{Program GO\,10265, PI.\,T.\,Brown.}.  

The photometric and astrometric reduction of ACS data, extensively described in
\cite{And08}, makes use of all the exposures simultaneously to generate a single
list of stars, which are measured independently in each image. The routines used
in the reduction exploit the independent dithered pointings of each image and
the knowledge of the point spread function (PSF) to detect a star and avoid
artifacts in the final star list. The reduction of the WFC3 images was performed
with a software (img2xym\_wfc3) that is mostly based on img2xym\_WFC
\citep{And06}. Star fluxes and positions were corrected for pixel area and
geometric distortions using the
solution given by \cite{Bel11} and finally calibrated into the VEGAmag as in
\citet{Bed05}. 

Lastly, we selected a high-quality sample of stars, i.e., relatively isolated
with small photometric, astrometric and PSF-fitting errors.  For this selection
we used the quality indices that our photometry software produces, in a
procedure that is described in detail by \citet{Mil09}.  By adopting the above
selection criteria, the $\sim$20\% of candidate cluster stars was rejected. The
sample of objects that was neglected from the analysis includes: i) stars that
are poorly measured due to contamination by cosmic rays or bad pixels; ii) stars
that are poorly measured due to contamination by nearby neighbor stars.
Moreover, we also neglected non-stellar objects (background galaxies), hot
pixels and artifacts associated to diffraction spikes of bright
stars~\citep[see\,][for details]{And08,Mil09,Mil12}.  For these objects we are
not able to properly estimate photometric and astrometric (random and
systematic) errors, because their profiles can strongly differ from the adopted
PSF model.

\begin{figure*}
\centering
\includegraphics[width=10cm]{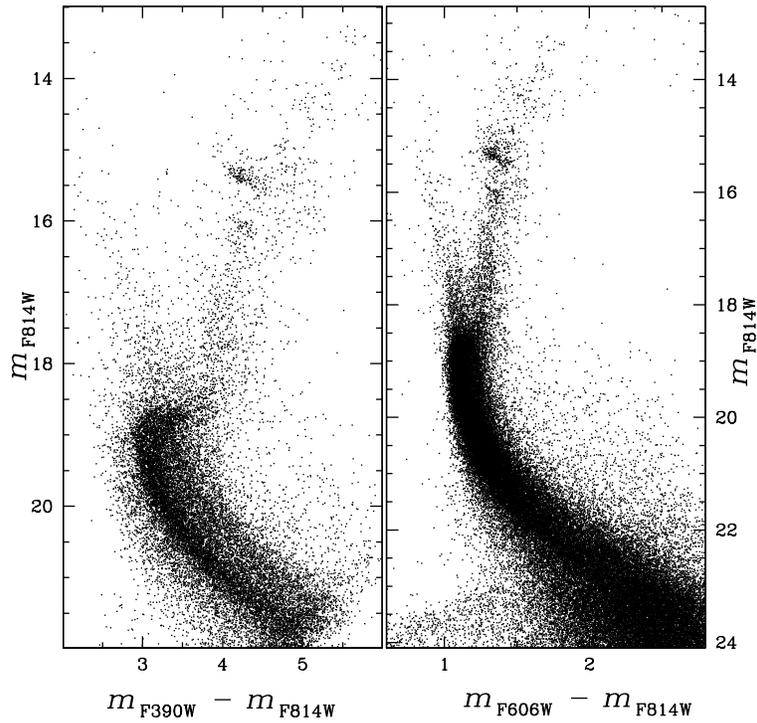}
\caption{Optical CMD of stars in the field of view of \ngc6528 based either on
WFC3/UVIS (\textit{left panel}) or on ACS/WFC photometry (\textit{right
panel}).}
\label{CMDs}
\end{figure*}

We ended up with an accurate photometric and astrometric catalog of $\sim
210,000$ stars in a magnitude interval ranging from nearly the tip of the RGB to
several magnitudes fainter than the Main Sequence Turn-Off (MSTO), reaching
$m_{\rm F814W} \approx 24$. The $m_{\rm F814W}$ versus $m_{\rm F390W}-m_{\rm F814W}$
CMD from WFC3/UVIS data and the  $m_{\rm F814W}$ versus $m_{\rm F606W}-m_{\rm
F814W}$ CMD from ACS/WFC data are shown in Fig.~\ref{CMDs} for stars in the
\ngc6528 FoV. 

\subsection{Proper Motions}\label{sec:PM} 
Proper motions (PMs) are obtained by comparing the position of stars measured at
two different epochs, following an approach that has been already used in
several papers \citep[e.g.\,][]{McL06,And03,Bed01}. In the present analysis the
average positions of the stars obtained by the optical ACS data and by both
optical and NIR WFC3 data, represent, respectively, the positions at the first
and second epoch. 

The computation of the geometric linear transformation between the coordinate
system of the first- and second-epoch catalogs is crucial for the determination
of the PM of the stars in our FoV. Therefore, we started off with the
identification of the same stars detected in both the epochs. Then, we picked
up a subsample made up only by the ones unsaturated and with a high signal-to-noise
ratio in all the ACS and WFC3 images. Moreover, since the cluster internal
motions are negligible compared to the field star motions, we eventually decided
to use cluster members as reference stars to define the transformation we are
looking for.

\begin{figure*}
\centering
\includegraphics[width=10cm]{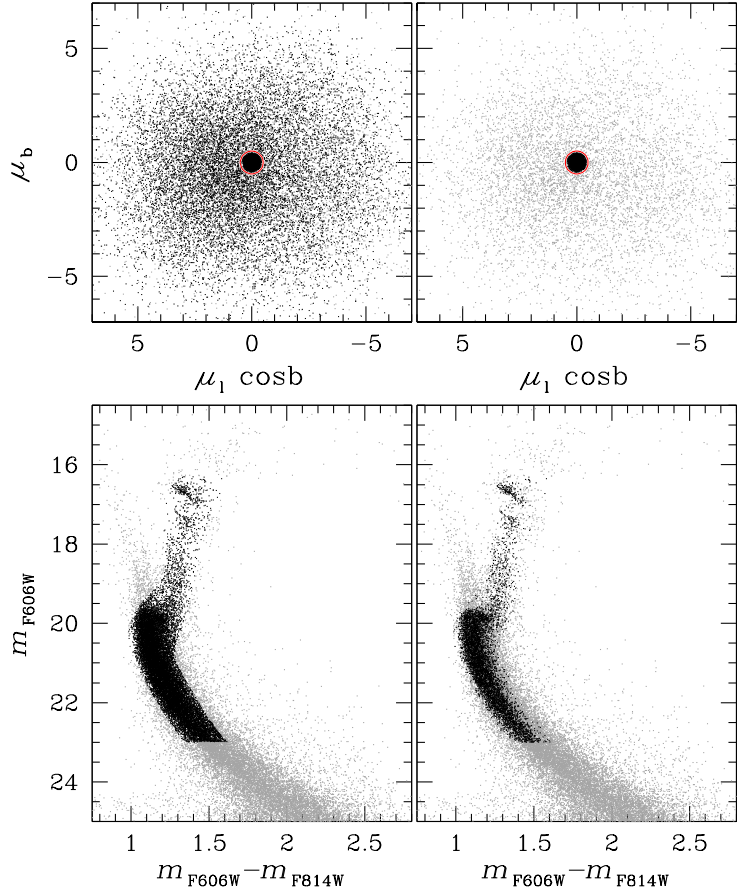}
\caption{Proper motion of stars in the field of view of \ngc6528. The stars
used as reference to measure proper motions are marked as black points in the
CMDs (bottom panels) and in the vector point diagram (top panels). The left
panels show the sample of reference stars selected only on the basis of their
position in the CMD and used to obtain a first preliminary estimate of proper
motions.  The right panels display the sample of reference stars used for the
final measurement of proper motions and selected on the basis of their position
both in the CMD and in the vector point diagram.}
\label{PMSEL}
\end{figure*}

In order to select the sample of cluster members as reference stars, we started
by  identifying all the  stars  that, on the basis of their position in the
CMD, are probable  cluster members, and obtained a raw proper-motion estimate by
using a local transformation based on this sample. As an example of the
criteria of selection we used, in the bottom-left panel of Fig.~\ref{PMSEL} we
plotted with black points all the stars that are used as reference to obtain the
first proper motion estimate. The same stars are plotted with the same color
code in the vector-point diagram (VPD) shown in the top-left panel. It is clear
that the PM of some reference stars is not consistent with their cluster
membership.

At this point, we drew the red circle which has a radius $3\,\sigma$, where
$\sigma$ is the average PM dispersion for reference stars in each direction and
contains the bulk of cluster stars.  We excluded   from the sample of reference
stars all  the stars that lie outside the red circle and do not have
cluster-like  proper motion even if they are placed near the  cluster sequence.
This  ensures that  the relative motion of the cluster stars is zero within  the
measurement errors.

\begin{figure*}
\centering
\includegraphics[width=10cm]{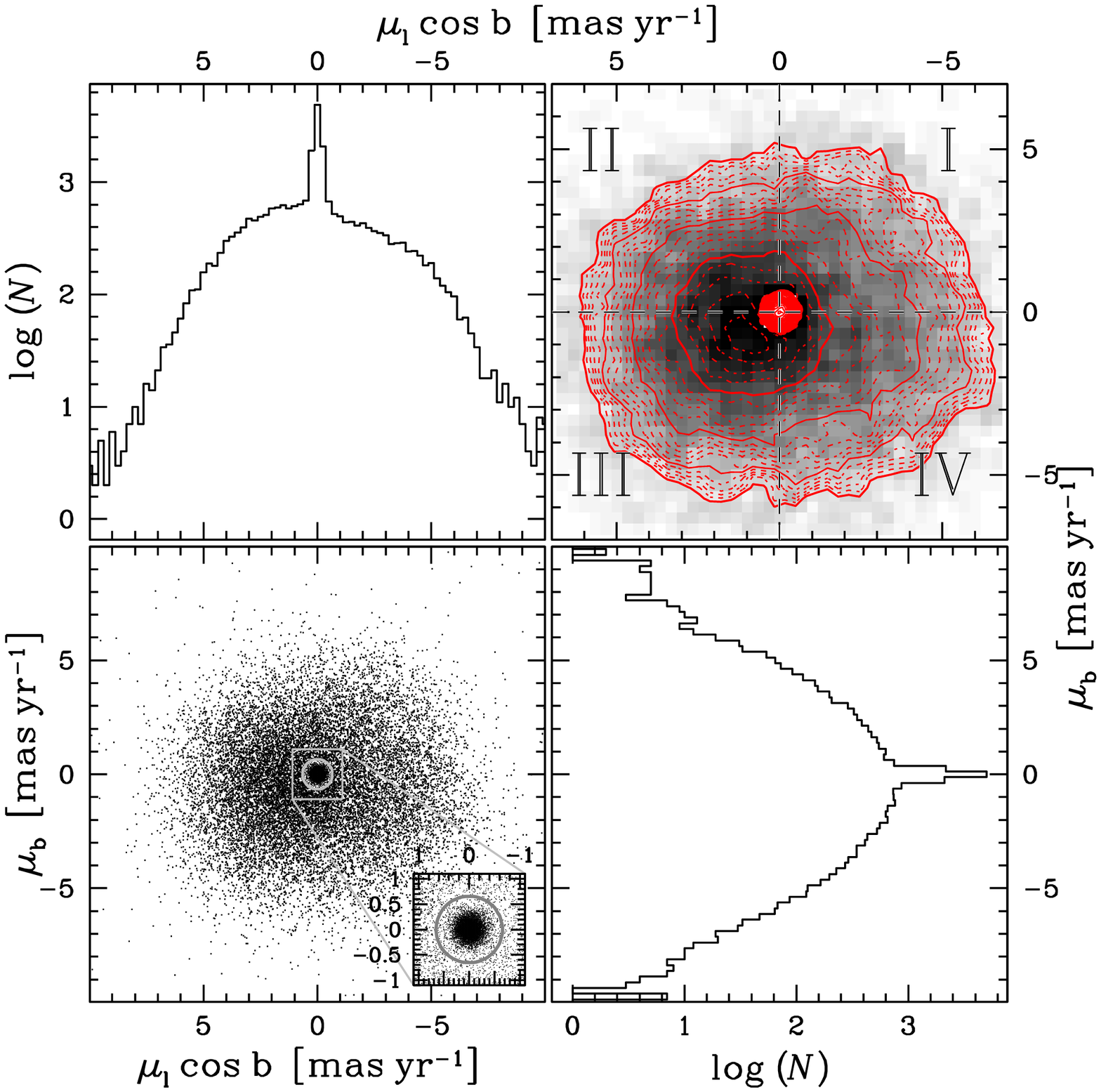}
\caption{
\textit{Bottom-left:} VPD of the proper-motion distribution of the entire sample
of stars: for the sake of clarity only stars with $m_{\rm F814W} < 22.1$ have
been plotted across the FoV of \ngc6528.  The inset shows a zoom-in around the
origin of the VPD. The bulk of stars around ($\mu_{\rm l} cos\,{\rm b}, \mu_{\rm
b}$) = (0, 0) mainly consists of cluster members.  \textit{Top-right:} Smoothed VPD
diagram and contours of the previous sample. The dashed lines separate the four
quadrants. The top-left and the bottom-right panels show the marginals of the
proper motions along the $\mu_{\rm l} cos\,{\rm b}$, and the $\mu_{\rm b}$ axis.}
\label{fig:pm} 
\end{figure*}

Results are shown in Fig.~\ref{fig:pm}. In the bottom-left panel we plotted the
VPD of the proper motions in Galactic coordinates for stars located across the
ACS FoV of \ngc6528: since proper motions were measured relative to a sample of
cluster members, the zero point of the motion is the mean motion of the cluster.
A zoomed VPD around the origin is shown in the inset. The bulk of stars
clustered around ($\mu_{\rm l} cos\,{\rm b}, \mu_{\rm b}$) = (0, 0) mainly includes
candidate cluster members while field stars are distributed over a broad range
of PMs. 

In the top-right panel of Fig.~\ref{fig:pm} a smoothed version of the VPD is
plotted, and the contours are superimposed onto the diagram.  The proper-motion
distribution of field stars is far from being symmetric across the four
quadrants. The left quadrants (II and III) show clear overdensities when
compared with the right quadrants (I and IV) and the density of the field stars
is maximum in the third quadrant as indicated by the contours. This evidence is
also supported by the marginals of the distribution drawn in the top-left
($\mu_{\rm l} cos\,{\rm b}$) and in the bottom-right ($\mu_{\rm b}$) panel. The above
distributions are skewed, and indeed the slopes at positive Galactic
longitudes and negative latitudes are steeper, respectively, than at
negative and positive ones. The mode of the field-star VPD also seems to be
displaced toward positive longitudes and negative latitudes, relative to the
cluster mode.
A further analysis of the VPD is provided in Sect.~\ref{Sect:Bulge}.

\subsection{Differential reddening}\label{sec:DR} 
The mean reddening in the direction of \ngc6528 is $E(B-V)\sim0.54$\,mag
\citep[][as updated in 2010]{Har96} and it is not uniform across the FoV.
Figure~\ref{CMDs} indeed shows that all the evolutionary sequences of \ngc6528
are broadened. The main culprit ought to be the differential reddening.

To minimize this effect on our analysis, we corrected our photometry for
differential reddening by using the method by \citet[][see their
Sect.~3.1]{Mil12}.  We note that the strong contamination from Bulge and Disk
stars that are present in the FoV of \ngc6528 can significantly affect results
of this procedure. To minimize the effect of field-stars contamination, we
determined the amount of differential reddening by using a sample of stars that
mainly including cluster members.  

\begin{figure*}
\centering
\includegraphics[width=10cm]{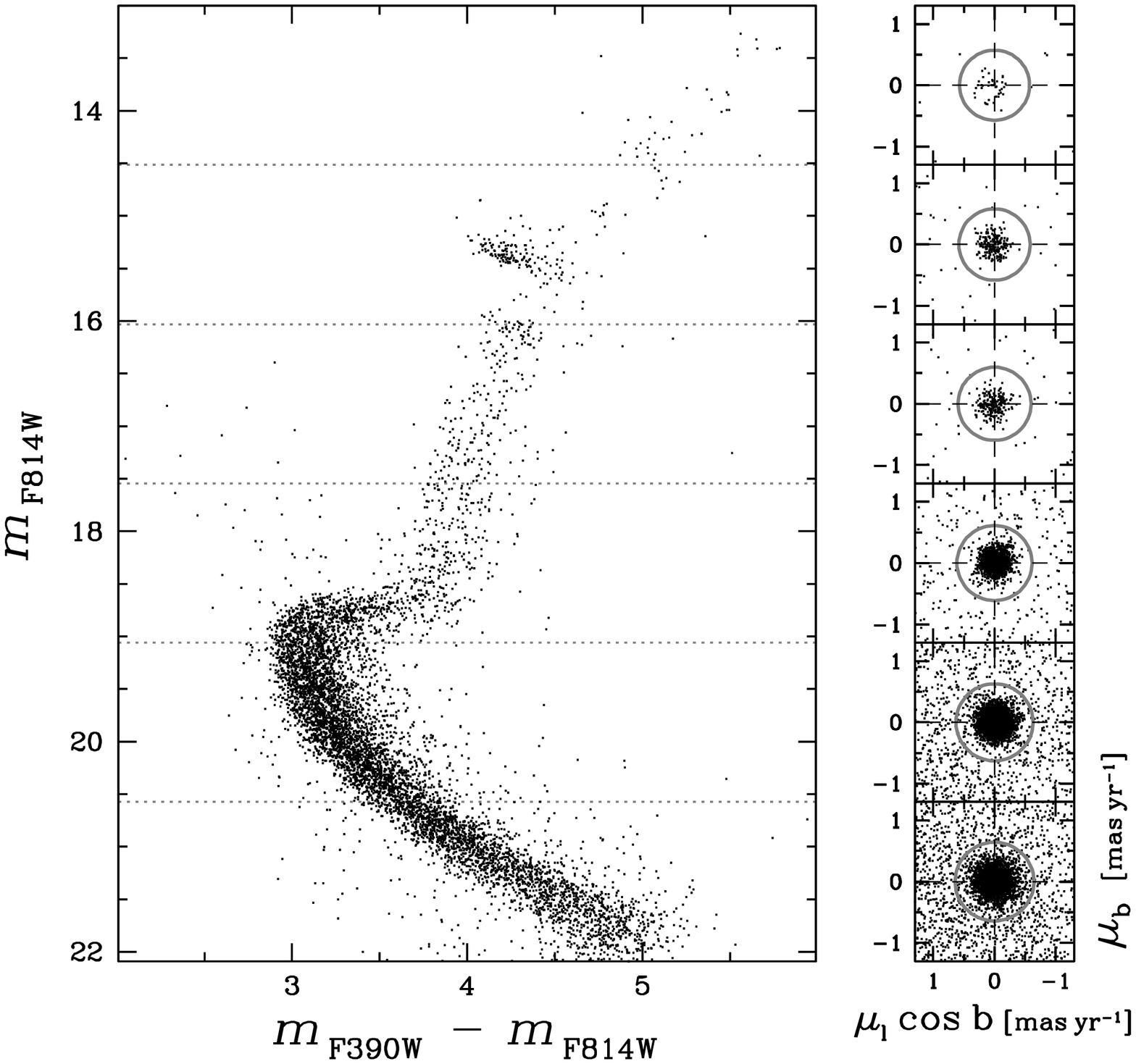}
\caption{ \textit{Left}: $m_{\rm F814W}, m_{\rm F390W} - m_{\rm F814W}$ CMD of
candidate cluster stars.  \textit{Right}: VPD for the stars in six magnitude
bins, marked by dotted lines in the left panel. Gray circle separate candidate
cluster members from field stars (see text for details).}
\label{fig:cm}
\end{figure*}

Candidate cluster members have been separated from field stars by following the
approach outlined in Fig.~\ref{fig:cm}.  The small boxes located in the right
panel display the VPD of the stars in the range $13 \lesssim m_{\rm F814W} \lesssim 22$,
divided into six bins of  about 1.5\,mag each (as indicated by horizontal dotted
lines in the left panel). For the sake of clarity, we zoomed the VPD around the
origin.

To identify the candidate cluster stars we firstly selected in each box of
Fig.~\ref{fig:cm} a sample of stars with proper  motion $\mu_{\rm
r}=\sqrt{(\mu_{\rm l}\,cos b)^{2}+\mu_{\rm b}^{2}}< 0.9$ mas\,yr$^{-1}$. This
sample mainly contains cluster members, but a mild contamination of field stars
is still present.

To minimize the field-star contamination we evaluated the dispersion of
proper-motion distribution in each panel ($\sigma(i)$ with $i =$1 to 6). Gray
circles have radius $r(i)=3\times\sigma(i)$ and we assumed cluster-members
candidates the stars located inside the individual circles.  Left panel of
Fig.~\ref{fig:cm} shows the $m_{\rm F814W}, m_{\rm F390W} - m_{\rm F814W}$ CMD
of the candidate cluster members.

The first step in the method used to correct for differential reddening is the
evaluation of the main-sequence fiducial line in the CMD. Then we selected, for
each star, the 45 nearest well-measured neighbors, in the catalog of candidate
cluster stars, and evaluated for each of them the color distance from the
fiducial line along the reddening line. We applied to the target star a
correction equal to the median color distance of these 45 stars \citep[details
of the procedure are given in][]{Mil12}. 

\begin{figure*}[!htp]
\centering
\includegraphics[width=10cm]{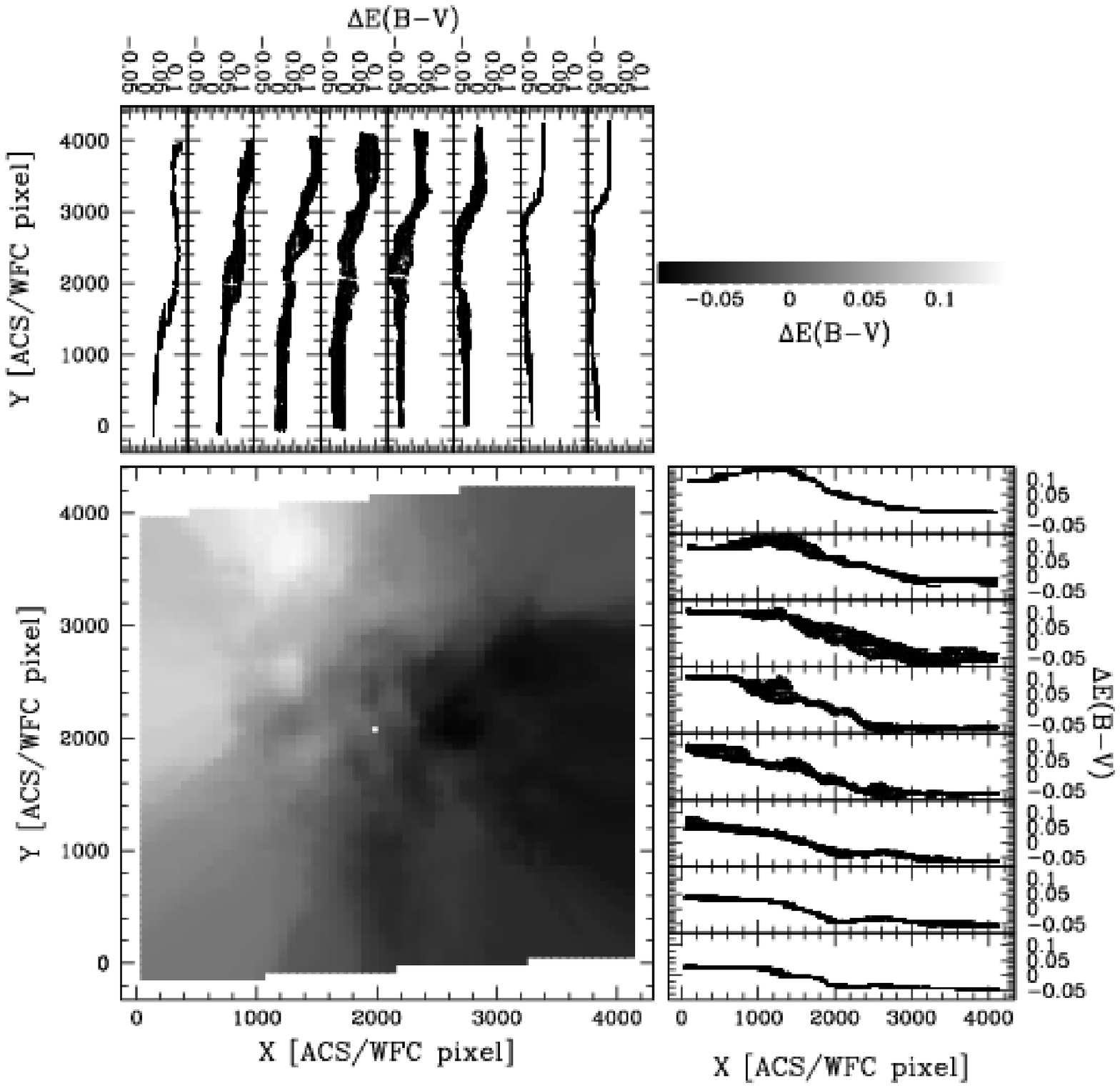}
\caption{\textit{Bottom-left:} Map of reddening across the FoV of \ngc6528.
Gray levels indicates different values of reddening as indicated on the top
right. The top and the bottom right panel display $\Delta\,E(B-V)$ as a function
of $Y$ and $X$, respectively, for stars located in eight vertical and horizontal
slices.}
\label{fig:redd}
\end{figure*}
 
\begin{figure*}
\begin{center}
\includegraphics[angle=0,width=10cm]{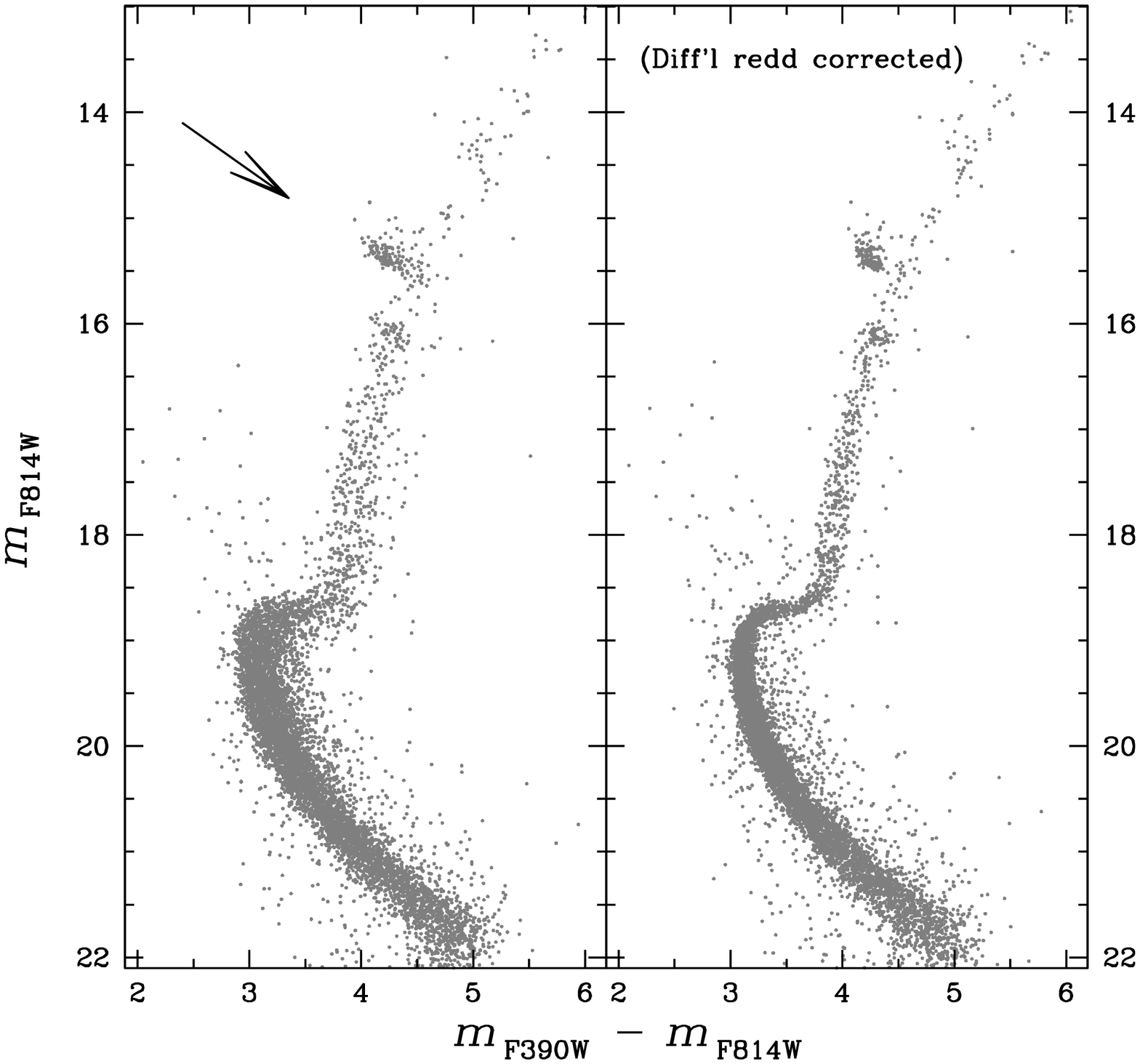}
\caption{$m_{\rm F814W}, m_{\rm F390W} - m_{\rm F814W}$ CMD of the candidate
cluster members before (left) and after the reddening correction (right).} 
\label{fig:3cmd}
\end{center}
\end{figure*}

The spatial reddening variation in the FoV of \ngc6528 is shown in
Fig.~\ref{fig:redd}.  Similarly to what done in \citet{Mil12}, we have divided
the FoV into eight horizontal slices and eight vertical slices and plot
$\Delta\,E(B-V)$ as a function of the $Y$ (top panels) and $X$ coordinate (right
panels). We have also divided the whole FoV into 64$\times$64 boxes of
128$\times$128 ACS/WFC pixels and calculated the average $\Delta~E(B-V)$ within
each of them.  The resulting reddening map is displayed in the bottom-left panel
where each box is represented as a gray square. The levels of gray are
indicative of the amount of differential reddening according to the scale on the
top right. Figure~\ref{fig:3cmd} compares the original CMD for cluster-members
candidates and the CMD corrected for differential reddening. 

\section{Selection of Bulge stars}\label{Sect:Bulge}
\begin{figure*} 
\centering 
\includegraphics[angle=0,width=10cm]{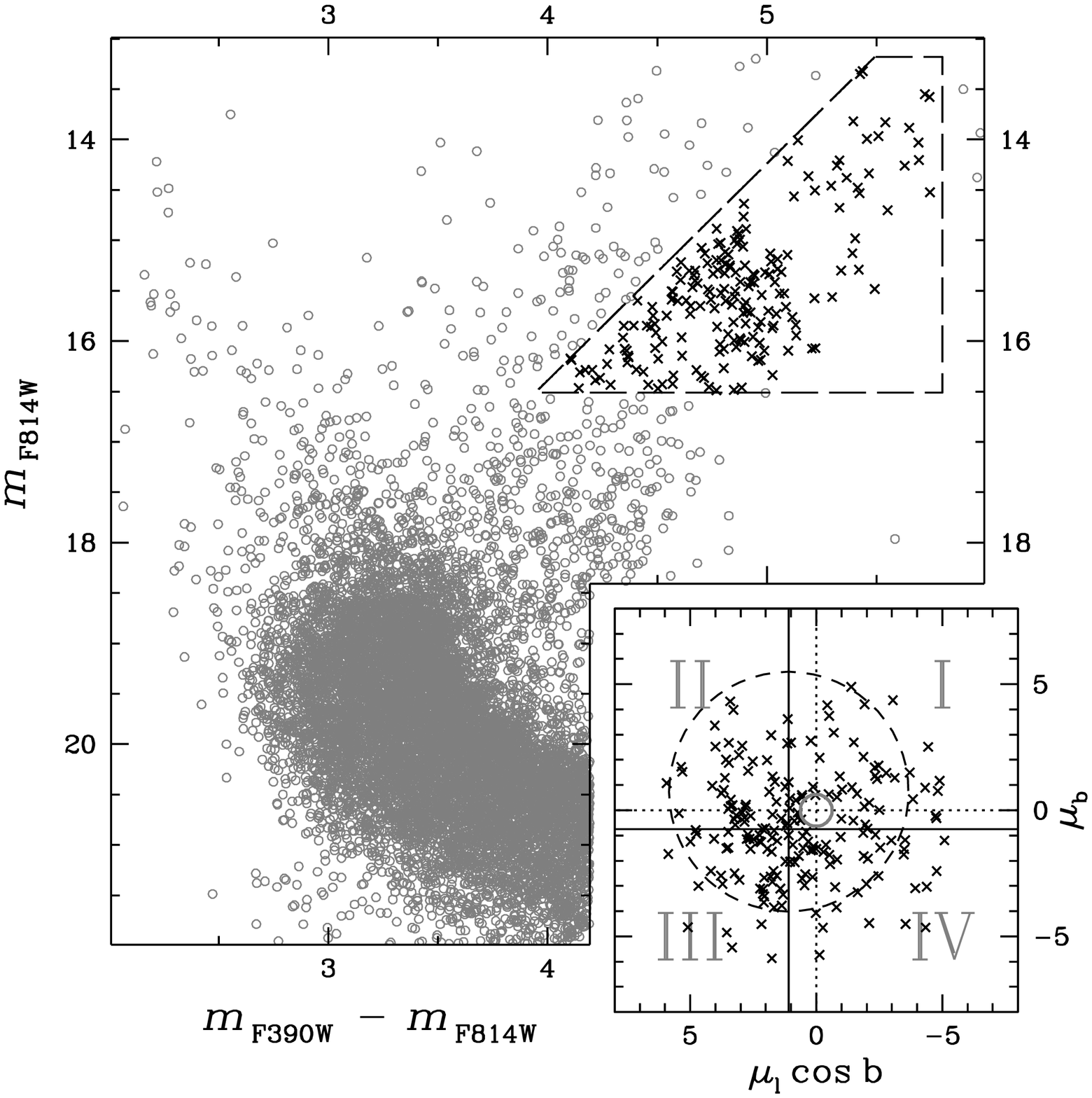}
\caption{$m_{\rm F814W}$, $m_{\rm F390W} - m_{\rm F814W}$ CMD of field stars.
The black crosses located inside the dashed window, mark stars that, according
to their position in the CMD, are probable Bulge members.  The position of these
stars in the VPD is shown in the inset.} 
\label{fig:selbulge1} 
\end{figure*}

In the previous section we have  described the procedure for the selection of
the candidate cluster members.  The $m_{\rm F814W}$,  $m_{\rm F390W}-m_{\rm
F814W}$ CMD of the remaining field stars is plotted in Fig.~\ref{fig:selbulge1}.
A glance at this CMD reveals a blue MS above the TO region that is populated by
young likely Disk stars, and a spread red RGB, that could be associated to the
Bulge.  

In order to select a sample of candidate Bulge members, among the field stars,
we started to estimate of the average motion of a small set of stars that, on the
basis of their position in the CMD, are probable Bulge members. In particular,
we picked up a subsample of the $\sim 200$ brightest and reddest RGB stars,
corresponding to the black crosses lying within the window outlined by the
dashed lines. 

The VPD of the selected stars has then been plotted (right-bottom panel) and the
median and the standard deviation of the PM distributions along the two Galactic
coordinates evaluated, obtaining: $\left<\mu_{\rm l} cos\,{\rm b}\right> = +1.09
\pm 0.2\,\rm mas\,yr^{-1}, \left<\mu_{\rm b}\right> = -0.74 \pm 0.2\,\rm
mas\,yr^{-1}; \sigma_{\mu_{\rm l} cos\,{\rm b}} = 2.57 \pm 0.2\,\rm
mas\,yr^{-1}, \sigma_{\mu_{\rm b}} = 2.18 \pm 0.16\,\rm mas\,yr^{-1}$.
Noticeably, the motion of the Bulge is centered on the quadrant III, thus
suggesting that the overdensity of stars observed in the VPD of
Fig.~\ref{fig:pm} and discussed in Sect.~\ref{sec:PM} is mainly due to the bulk
motion of Bulge stars. Our estimates of the Bulge velocity dispersion are
consistent with those provided by \citet{Fel02}, who found $\sigma_{\mu_{\rm l}
cos\,{\rm b}} = 3.27 \pm 0.27\,\rm mas\,yr^{-1}, \sigma_{\mu_{\rm b}} = 2.54 \pm
0.17\,\rm mas\,yr^{-1}$ and \citet{Kui02}, who found $\sigma_{\rm \mu_{\rm l}
cos\,{\rm b}} = 2.98 \pm 0.017$ and $\sigma_{\mu_{\rm b}} = 2.54 \pm 0.014\,\rm
mas\,yr^{-1}$ for stars in the Baade's Window.

\begin{figure*}
\centering
\includegraphics[angle=0,width=10cm]{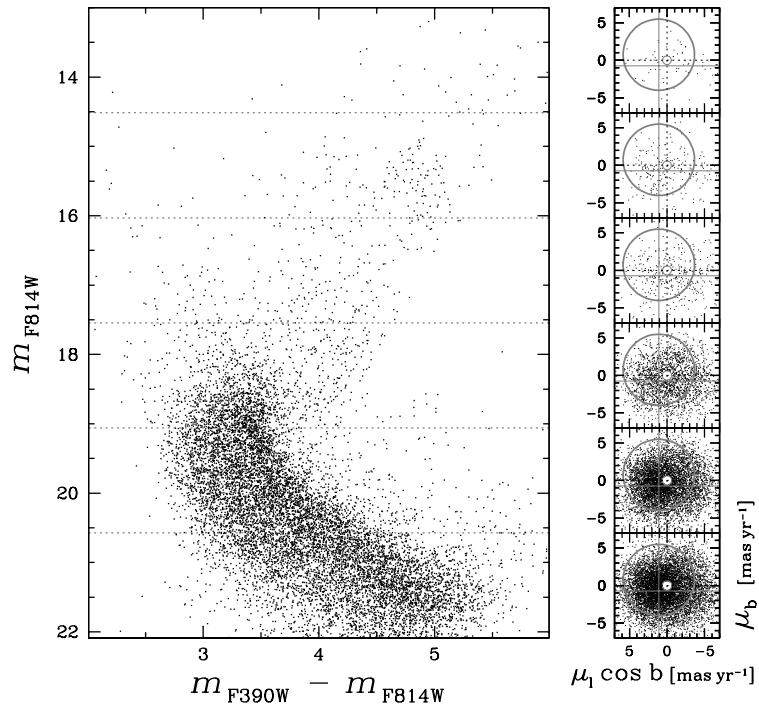}
\caption{\textit{Left panel:} $m_{\rm F814W}$ versus $m_{\rm F390W}-m_{\rm
F814W}$ CMD for candidate Bulge stars. \textit{Right panels:} VPD of field stars
in six magnitude bins. Gray circles separate candidate Bulge stars from
remaining field stars. See text for details.} 
\label{fig:selbulge2}
\end{figure*}

The Bulge membership of the stars in our catalog has then been established
following an approach similar to that used to select the candidates of cluster
population, as displayed in the right panels of Fig.~\ref{fig:selbulge2}. All
the stars not belonging to \ngc6528, whose PMs are located inside a circle with
a radius equal to $2 \times \sigma_{\rm b}$ where $\sigma_{\rm b}=2.38$ is the
average value of the standard deviation along the two Galactic coordinates, are
flagged as possible Bulge members.  The $m_{\rm F814W}$ versus $m_{\rm
F390W}-m_{\rm F814W}$ CMD of selected Bulge stars is plotted in the left panel
of Fig.~\ref{fig:selbulge2}. 

\section{The CMDs of \ngc6528 and Bulge stars}
\begin{figure*}
\centering
\includegraphics[width=10cm]{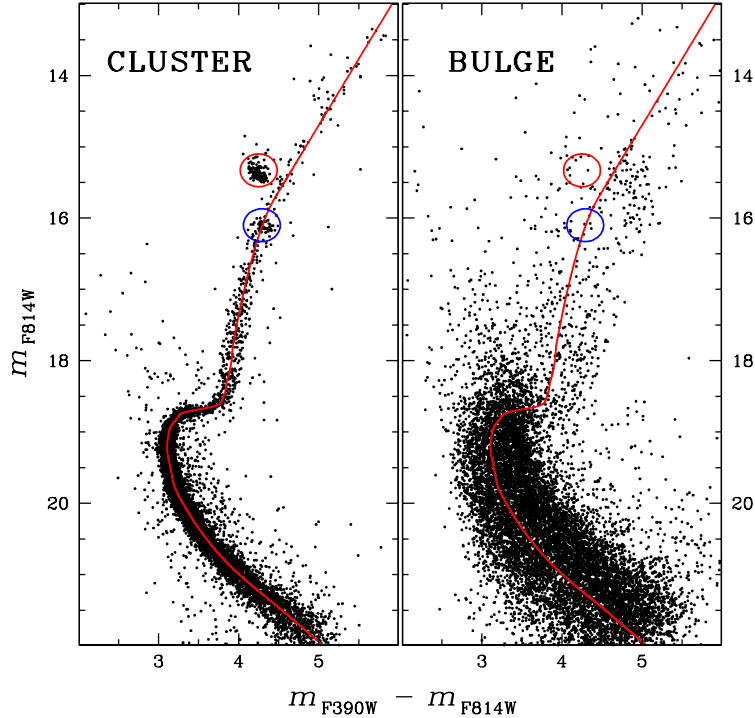}
\caption{\textit{Left panel:} $m_{\rm F814W}$ versus $m_{\rm F390W}-m_{\rm
F814W}$ CMD of candidate cluster (\ngc6528) members. The red solid line shows
the ridge line of \ngc6528, while the red and the blue ellipses mark the
position of red HB stars and of RGB bump stars. \textit{Right panel:} Same as
the left, but for Bulge stars. The line and the symbols are the same of the
left panel.}\label{fig:CMDs}
\end{figure*}

In this section we analyze the CMDs of candidate cluster and Bulge stars. The
left panel of Fig.~\ref{fig:CMDs} shows the $m_{\rm F814W}$, $m_{\rm
F390W} - m_{\rm F814W}$ CMD of \ngc6528. The selection of candidate cluster
members based on proper motions and, the correction for differential reddening
improved the photometric precision and allowed us to disclose several
interesting evolutionary features of \ngc6528.

i) The HB morphology is typical of metal-rich GGCs, and indeed central
He-burning stars are located in a well defined red HB (red ellipse;
$\left<m_{\rm F814W}\right> = 15.33, \left<m_{\rm F390W} - m_{\rm F814W}\right>
= 4.25$\,mag). 

ii) Stars located across the RGB bump (blue ellipse; $\left<m_{\rm F814W} =
16.10\right>, \left<m_{\rm F390W} - m_{\rm F814W}\right> = 4.29$\,mag) display a
well defined overdensity.  The luminosity  --- as expected in metal-rich GGCs
--- is fainter than the red HB.

iii) Evolved (SGB, RGB) and unevolved (MS) evolutionary features display a tight
distribution in color. Indeed, the spread in color around the cluster ridge line
(red solid line) is typically smaller than 0.04 mag. 

The CMD of candidate Bulge is plotted in the right panel of Fig.~\ref{fig:CMDs}.
The comparison with the cluster ridge line indicates that the bulk of Bulge red
giants (RGs) are systematically redder than cluster RGs, thus suggesting that
they are --- on average more --- metal-rich. This working hypothesis is also
supported by the fact that the SGB of Bulge stars ($m_{\rm F814W} \simeq 18.75,
3.6\lesssim m_{\rm F390W} - m_{\rm F814W} \lesssim 4$) shows a flatter color
distribution than cluster SGB stars.  Moreover, Bulge RGs show, at fixed
luminosity, a larger spread in color when compared with cluster RGs. These
empirical evidence suggests that Bulge RGs are affected either by a more complex
reddening distribution along the line of sight and/or by a larger spread in
metallicity.  

The MS of Bulge stars is affected by a large spread in color, but the morphology
of the MS across the TO region is quite similar to the cluster ridge line. The
same outcome applies to red HB stars, and indeed they cover a broader range both
in magnitude ($m_{\rm F814W} \approx 15.2 \div 16.2$) and in color ($m_{\rm
F390W} - m_{\rm F814W} \approx 4.7 \div 5.1$). Moreover, the blue MS ($m_{\rm
F814W} \le 18.5, m_{\rm F390W} - m_{\rm F814W} \le 3.40$) tracing a young Disk
population seems poorly populated. The above evidence confirms that residual
contamination from the Disk is negligible for our purposes and the field stars
we are dealing with are mainly Bulge stars.  Moreover, the latter ones appear to
be coeval with candidate cluster stars, thus supporting earlier findings by
\citet{Ort01}.  

More quantitative estimates of the Bulge stellar population(s) are however
hampered by the notorious degeneracies among differential reddening, metallicity
and stellar ages that are typical of optical CMDs.    


\section{Empirical constraints}
To constrain the evolutionary properties of cluster and field stars, we adopted
two empirical calibrators. We selected the old, metal-rich GGC 47\,Tucan\ae\ =
\ngc104 \citep[$t\approx11\,{\rm Gyr}, \feh = -0.8$;][]{Van10,Car09} and the
old, super-metal-rich open cluster \ngc6791 \citep[$t\approx8\,{\rm Gyr},\feh =
+0.3$;][]{Bra10,Boe09}. For 47\,Tuc we used accurate ACS/WFC photometry in
$F606W$ and $F814W$ bands from \citet{And08} and \citet{Cal12}.  The photometry of
\ngc6791, has been performed for this investigation (see
Sect.~\ref{sec:obs})\footnote{Note that the above clusters were selected because
our group already performed accurate WFC3/IR photometry \citep{Cal12}. This
means that we are following the very same approach to perform the photometry and
to derive the cluster ridge lines.  Moreover, we plan to use the same empirical
calibrators for both optical and NIR photometry.}. 

\begin{figure*}
\centering
\includegraphics[width=10cm]{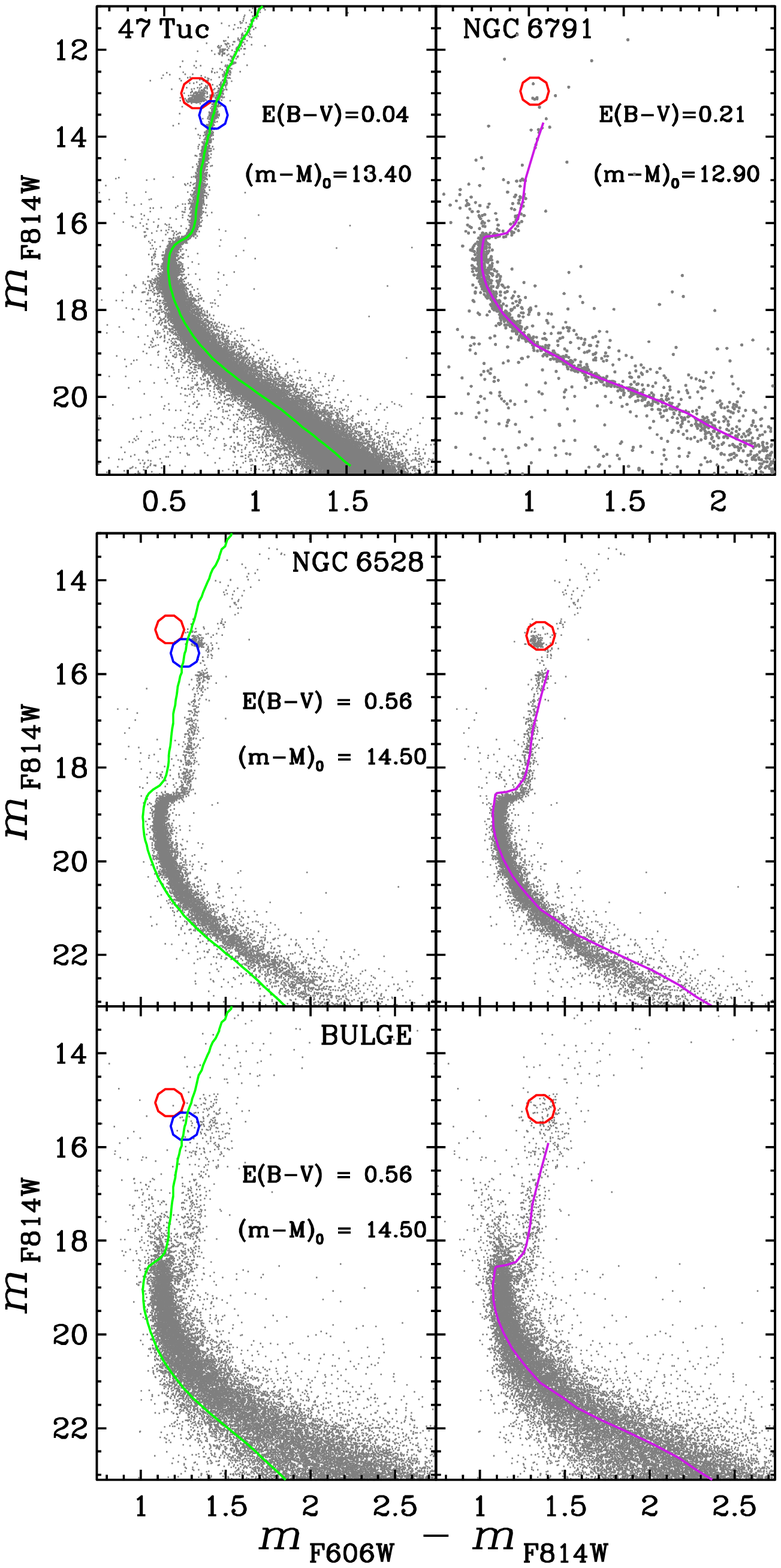}
\caption{\textit{Top-left}: $m_{\rm F814W}, m_{\rm F606W} - m_{\rm F814W}$
CMD of 47\,Tuc based on images collected with ACS at HST.  The red and blue
circles mark,respectively, the red HB and the RGB bump, while the green line the cluster
ridge line.  \textit{Top-right}: Same as the left, but for the old, metal-rich
cluster \ngc6791. The red circle marks the position of red clump stars.
\textit{Middle-left}: Comparison between the $m_{\rm F814W}, m_{\rm F606W} -
m_{\rm F814W}$ CMD of \ngc6528 and the ridge line of 47\,Tuc. The adopted true
distance modulus and reddening (\citeauthor{Har96} \citeyear{Har96}) are
labeled. \textit{Middle-right}:  Same as the left, but the comparison is with
the old, metal-rich cluster \ngc6791. \textit{Bottom-left}: Same as the middle,
but the comparison is between candidate Bulge stars and the ridge line of
47\,Tuc. \textit{Bottom-right}: Same as the left, but the comparison is with the
old, metal-rich cluster \ngc6791.}
\label{fig:calib} 
\end{figure*}

Data plotted in the top panels of Fig.~\ref{fig:calib} display the $m_{\rm
F814W}, m_{\rm F606W} - m_{\rm F814W}$ CMD of 47\,Tuc (left) and of \ngc6791
(right). The red and blue circles in the left panel identify, respectively, the
red HB ($\left<m_{\rm F814W}\right>=13.00, \left<m_{\rm F606W} - m_{\rm
F814W}\right>=0.68$\,mag) and the  RGB bump ($\left<m_{\rm F814W}\right> = 13.5,
\left<m_{\rm F606W} - m_{\rm F814W}\right> = 0.77$\,mag), while the green curve
shows the ridge line of the cluster. Symbols in the right panel are the same as
in the left panel, but the red circle centered on $m_{\rm F814W} = 13.2, m_{\rm
F606W} - m_{\rm F814W} = 1.03$\,mag marks red clump stars (the ridge line is
shown in magenta). The RGB bump was not identified, since the number of stars
along the RGB in this cluster is quite limited.     

To compare the different clusters we adopted for \ngc6528 a true distance
modulus of 14.50 and a cluster reddening of $E(B-V)=0.56$\,mag \citep{Har96}.
These values are consistent with the estimates available in the literature
($\mu_0 = 15.10, E(B-V) = 0.46$, \citeauthor{Zoc04} \citeyear{Zoc04}; $\mu_0 =
14.49, E(B-V)=0.54$, \citeauthor{Mom03} \citeyear{Mom03}). The reddening was
transformed into the ACS bands by interpolating the extinction values provided
in \citet{Bed05}, and we found $E(F606W-F814W) \simeq 0.53$\,mag.     

The comparison between candidate members of \ngc6528 and the fiducial line of
47\,Tuc (middle left panel) seems to indicate that the former is more metal-rich
than the latter.  The hypothesis that the difference is mainly in chemical
composition and not in cluster age is supported by the following evidence:  

i) the entire MS in 47\,Tuc is systematically bluer than in \ngc6528; 

ii) the RGB in \ngc6528 has a shallower slope than the RGB of 47\,Tuc; 

iii) both the red HB and the RGB bump of 47\,Tuc are brighter and bluer 
than those in \ngc6528. 

On the other hand, the comparison between \ngc6528 and \ngc6791 indicates that
the former cluster is less metal-rich than the latter. The hypothesis that the
difference is in both chemical composition and age is supported by the following
evidence: 

i) the ridge line of \ngc6791 becomes, for magnitudes fainter than $m_{\rm
F814W} \sim 21.5$, systematically redder than MS stars in \ngc6528.  Moreover,
the ridge line of \ngc6791 attains colors in the TO region that are
systematically brighter than MSTO stars in \ngc6528;  

ii) the shape and the extent in color of the sub-giant branch region in \ngc6791
is narrower ($m_{\rm F606W} - m_{\rm F814W} \sim 0.85$ {\it vs} $m_{F606W} -
m_{\rm F814W} \sim 1.2$) compared with \ngc6528.  

The comparison with the candidate Bulge stars does not permit firm constraints
on their evolutionary properties. However, field RGs are systematically redder
and display a shallower slope when compared with  the 47\,Tuc ridge line.
Moreover, the distribution in magnitude and in color of field SGB stars are
narrower than in 47\,Tuc.  The above evidence indicates that Bulge stars are
also more metal-rich than 47\,Tuc (bottom-left panel of Fig.~\ref{fig:calib}).   

Data plotted in the bottom right panel of Fig.~\ref{fig:calib} show  a plausible
agreement with the ridge line of  \ngc6791, and indeed the slope of the cluster
RGB is quite similar to Bulge RG stars. Moreover, field MSTO stars attain colors
($m_{\rm F606W} - m_{\rm F814W} \sim 1.1$) and magnitudes ($m_{\rm F814W} \sim
18.5 \div 19.5$) similar to the ridge line of \ngc6791. The above evidence
indicates that Bulge stars might have a spread in age of the order of a few
Gyrs. Current findings rely on a very limited fraction of Bulge stars, more
quantitative constraints do require accurate and deep photometry and proper
motion estimates along different line of sights of both the inner and the outer
bulge.  

\begin{figure*}
\centering
\includegraphics[width=10cm]{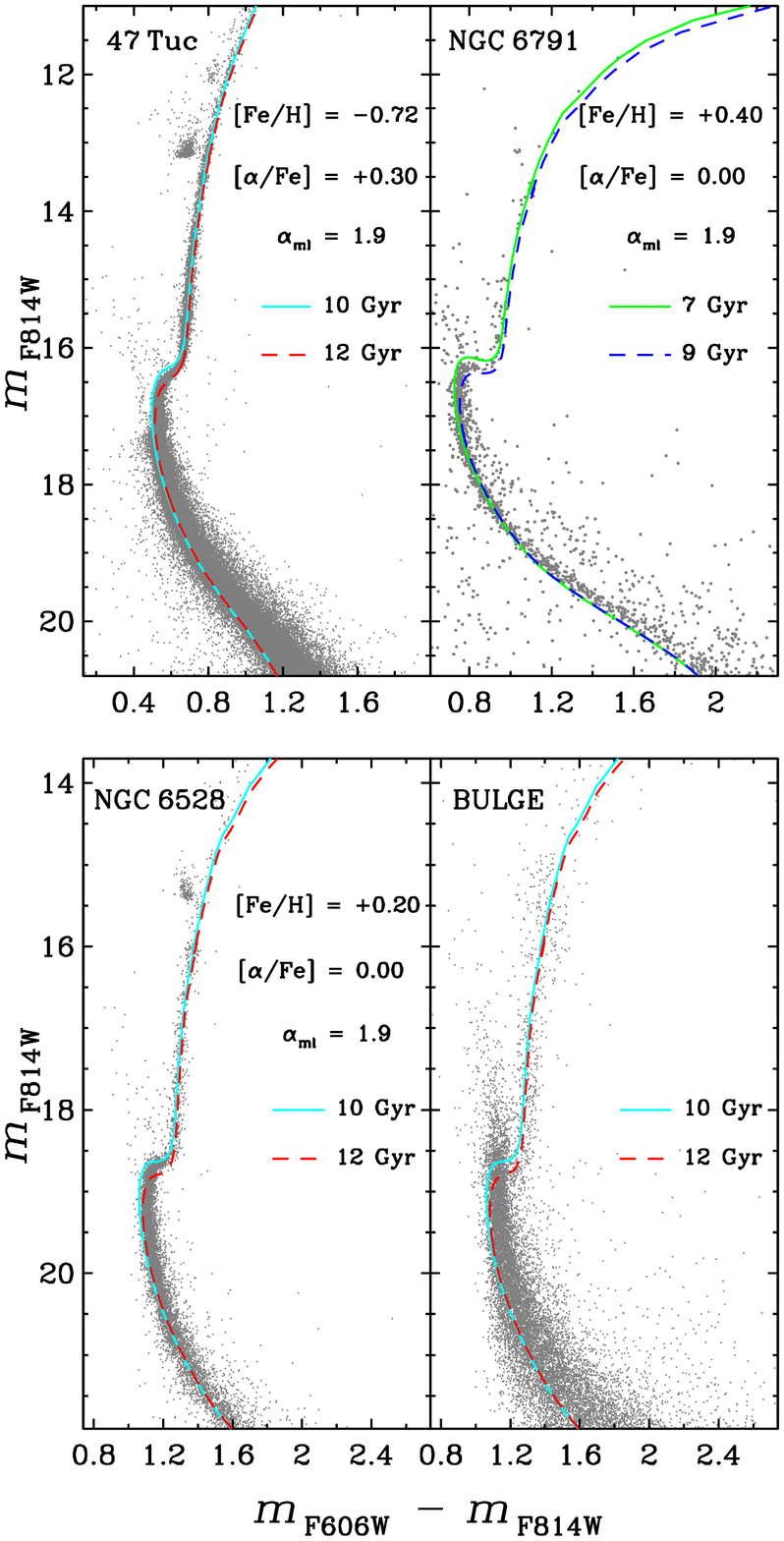}
\caption{\textit{Top-left}: Comparison in the $m_{\rm F814W}, m_{\rm F606W}
- m_{\rm F814W}$ CMD between stellar isochrones and optical ACS photometry for
the GGC 47\,Tuc. The cluster isochrones range from 10 to 12\,Gyr and are based
on a set of $\alpha$-enhanced evolutionary models constructed at fixed chemical
composition (see labeled values). The adopted true distance modulus and
reddening are the same as in Fig.~\ref{fig:calib}. \textit{Top-right}: Same as
the left, but for the cluster \ngc6791. The isochrones range from 7 to 9\,Gyr.
They were constructed by assuming a scaled-solar mixture and a super-metal-rich
chemical composition.  \textit{Bottom-left}: Same as the top, but for the GGC
\ngc6528. The cluster isochrones range from 10 to 12\,Gyr. They were constructed
assuming a scaled-solar mixture and a super-metal-rich chemical composition.
\textit{Bottom-right}: Same as the left, but for the candidate Bulge stars.
\label{fig:iso}}  
\end{figure*}


\section{Theoretical constraints}
We computed stellar tracks and isochrones with metallicities Z=0.004, 0.010,
0.015, 0.020, 0.025, 0.035. The initial helium abundance was fixed by assuming a
linear helium-to-metal enrichment ratio: $Y=Y_{\rm p}+\Delta Y/\Delta Z\,Z$,
with a cosmological He abundance $Y_{\rm p}=0.2485$
\citep{steigman06,peimbert07} and $\Delta Y/\Delta Z=2$
\citep{pagel98,jimenez03,gennaro10}. 

The different sets of evolutionary models were constructed assuming the
following chemical compositions: ($Z$, $Y$): 0.004, 0.256; 0.010, 0.269; 0.015,
0.278; 0.020, 0.288; 0.025, 0.299; 0.030, 0.308. We adopted both scaled-solar
\citet{asplund09} and $\alpha$-enhanced ($\afe = +0.3$) chemical mixtures.  For
each set, the evolutionary models were computed for masses ranging from 0.30 to
1.20 \msun ($\Delta\,M = 0.05$ \msun). Three different values of the mixing
length parameter --- $\alpha_{\rm ml} = 1.7, 1.8, 1.9$ --- were adopted.  This large
grid of models was computed with the FRANEC evolutionary code (Pisa Stellar
Evolution Data Base, \citeauthor{database2012} \citeyear{database2012};
\citeauthor{deglinnocenti08} \citeyear{deglinnocenti08}). The stellar isochrones
range from 5 to 15\,Gyr. To transform evolutionary prescriptions into the
observational plane, we adopted the synthetic spectra by \cite{brott05} for
$T_{\rm eff}\le10000$\,K and by \citet{castelli03} for $T_{\rm eff}>10000$\,K.
The synthetic spectra were computed assuming a scaled-solar chemical mixture and
have the same global metallicity of evolutionary models.

To validate the cluster isochrones, we estimated the age of the two calibrating
clusters. Top panels of Fig.~\ref{fig:iso} show the comparison in the $m_{\rm
F814W}, m_{\rm F606W} - {\rm F814W}$ CMD between isochrones and observations for
47\,Tuc (right) and \ngc6791 (left).  The isochrones that provide the best fit
with 47\,Tuc data are based on evolutionary models constructed assuming a mixing
length $\alpha_{\rm ml}=1.9$, an $\alpha$-enhanced ($ \afe=+0.30$) chemical
mixture and a metal-rich ($ \feh=-0.72$) composition.  We assumed for 47\,Tuc
the distance modulus and the reddening labeled in Fig.~\ref{fig:calib}. The
comparison between isochrones and observations shown in Fig.~\ref{fig:iso}
suggest that the absolute age of 47\,Tuc is $\sim$$11\pm1$\,Gyr, in agreement
with previous estimates from different authors\footnote{The literature
concerning distance modulus, reddening and age estimates of 47\,Tuc is quite
broad: $(m-M)_0 = 13.18\pm0.07$, $E(B-V) = 0.04\pm0.02$, $t\approx11.5$\,Gyr,
\citep{Sal07}; $(m-M)_{\rm V} \simeq 13.375$, $E(B-V) \simeq 0.04$,
$t\approx12$\,Gyr, \citep{Ber09}; $(m-M)_{\rm V} \simeq 13.40$, $E(B-V)\simeq
0.032$, $t\approx11$\,Gyr, \citep{Van10}; $(m-M)_{\rm V}= 13.30, E(B-V)= 0.03$,
$t=12.75\pm0.5$\,Gyr \citep{Dot10}, $t=11.75\pm0.25$\,Gyr \citep{Van13}; and
with \citet{Mar09}: 12.3--13.7\,Gyr, using different sets of isochrones.  More
recently, \citet{han13} have inferred a younger age of $t=9.9\pm0.7$\,Gyr from
their analysis of the white dwarf cooling sequence.} 

The isochrones to constrain the age of the open cluster \ngc6791 were
constructed by assuming a scaled-solar ($\afe=0$) chemical mixture and a
super-metal-rich ($\feh=0.40$) composition. We used the same distance modulus
and reddening labeled in Fig.~\ref{fig:calib} and infer that the absolute age of
\ngc6791 is $\sim 8\pm1$\,Gyr in agreement with similar literature
estimates\footnote{The literature distance modulus, reddening and age estimates
of \ngc6791 cover a broad range: $(m-M)_0\sim13.57$, $E(B-V)\sim0.15$,
$t\approx8$\,Gyr, \citep{Bra10}; $(m-M)_{\rm V}\sim13.51$, $E(B-V)\sim0.14$,
$t\approx8.3$\,Gyr, \citep{Bro12}}. 

To constrain the absolute age of \ngc6528 we performed a series of fits using
different chemical mixtures and different iron abundances. The bottom-left panel
of Fig.~\ref{fig:iso} shows the comparison of the observed $F814W, F606W-F814W$
CMD with best-fit isochrones.  Data plotted in this panel indicate that \ngc6528
seems to be coeval with 47\,Tuc, and indeed we found that its age is
$11\pm1$\,Gyr.  Moreover, the shape of both the SGB and the RGB support the
hypothesis that this globular has a super-metal-rich iron abundance.  The
cluster age agrees quite well with the most recent estimates available in the
literature: $t \sim 13$ Gyr, \citep{Zoc01}, $t=11 \pm 2$, \citep{Fel02};
$t\sim12.6$\,Gyr, \citep{Mom03}; $t\sim12.5$\,Gyr, \citep{Bro05}.

The bottom-right panel shows the comparison between the candidate Bulge stars
and the same set of old isochrones. A detailed analysis of the age and
metallicity distribution of Bulge stars is beyond the aims of the current
investigation. However, data plotted in the bottom-right panel display that old,
super metal-rich isochrones account for a sizable sample of field RGs and a
plausible fit for the bulk of SGB and MSTO stars. The comparison further
supports the evidence for a spread both in age and in chemical composition.    

We note that the adopted iron abundance for \ngc6528, $\feh = +0.2$ is slightly
higher than current spectroscopic measurements ($\feh = -0.1 \pm 0.2$,
\citeauthor{Zoc04} \citeyear{Zoc04}; $\feh = -0.17 \pm 0.01$ \citeauthor{Ori05}
\citeyear{Ori05}; $\feh = +0.07 \pm 0.08$ \citeauthor{Car01} \citeyear{Car01}).
No firm conclusion can be reached concerning the abundance of $\alpha$ elements.
The fit with scaled-solar isochrones is slightly better than with
$\alpha$-enhanced isochrones. The same outcome applies to spectroscopic
observations, and indeed, \cite{Zoc04} found $\afe\approx+0.1$, while
\cite{Ori05} and \cite{Car01} found $\afe\approx+0.3$. Independent constraints
will require accurate and deep NIR photometry and larger samples of
high-resolution spectra.  


\section{Summary and final remarks}
On the basis of a large set of optical and near-infrared images collected with
ACS/WFC, WFC3/UVIS, and WFC3/IR on board the HST, we performed deep and precise
photometry of the Galactic globular \ngc6528. Current data set covers a time
interval of almost ten years and allowed us to perform accurate proper motion
measurements over the entire FoV of the above images. The proper
motions were used to separate candidate field and cluster stars.  

Moreover, we also performed detailed estimates of the reddening map over the
entire field of view of the above images and provided reddening corrections for
individual objects. The proper motion selection and the correction for
differential reddening brought forward very accurate and deep CMDs for both
field and cluster stars. This applies not only to evolved evolutionary features
(red horizontal branch, red clump, RGB bump), but also to MS stars fainter than
the MSTO.     

The comparison with empirical calibrators (47\,Tuc, \ngc6791) and with cluster
isochrones indicate that \ngc6528 is an old GGC with super-solar iron abundance.
Together with 47\,Tuc, it seems to belong to the group of old GGCs that are
characterized by a minimal dispersion in age and no evidence of an
age-metallicity relation \cite{Mar09}.  The above findings together with the
solid empirical evidence of an overabundance of $\alpha$-elements in field Bulge
stars over the entire metallicity range \citep{mcw94,Zoc06,gon11,nes13a,nes13b}
further support the hypothesis that the high rate of star formation in the
innermost regions of the Bulge caused, in the first $\sim$ 1\,Gyr, a fast
chemical enrichment of the interstellar medium.

Our findings concerning \ngc6528, support previous results from different
authors on a rapid chemical enrichment of the Bulge.  In particular, a thorough
high-spectral-resolution spectroscopic survey of the Galactic Bulge (ARGOS) was
recently performed by \citet{fre13} and \citet{nes13a,nes13b}. On the basis of
the metallicity distribution function and of the kinematic properties of a large
sample of RC stars they found that the metal-rich stellar component ([Fe/H]
$>-0.5$) is associated with the boxy/peanut \citep{McW10} Bulge (components A,
B), while the metal-poor component ([Fe/H] $< -0.5$) is associated with both
the Galactic metal-intermediate thick disc and the metal-poor halo (components
C, D, E).  The Bulge stellar components include two spatially and chemically
distinct sub-populations: i) A thin more metal-rich sub-sample ([Fe/H]
$\approx$ 0.15) that is kinematically colder and closer to the Galactic plane.
This component is associated to the thin disc of the Galaxy (A component).  ii)
A thick, more metal-poor sub-sample ([Fe/H] $\approx -0.25$) that is uniformly
distributed across the selected fields. This component is associated to the
boxy/peanut Bulge and it is considered the tracer of the pristine Bulge stellar
population. It is also characterized by a large vertical distribution formed via
an early disc instability that dragged the first generation of the thin disc
stars into the boxy/peanut Bulge structure \citep{nes13a,nes13b}.

It goes without saying that current empirical scenario concerning the
metallicity distribution of Bulge stars awaits for a comprehensive analysis of
the iron distribution and of the $\alpha$-element distribution of stellar
tracers tightly connected with the pristine stellar population of the Bulge,
namely blue/red horizontal branch stars and RR\,Lyr\ae\ stars.   


\begin{acknowledgements}
It is a pleasure to thank an anonymous referee for his/her pertinent suggestions
and comments that improved the content and the readability of the manuscript.
This work was partially supported by PRIN INAF 2011 ``Tracing the formation and
evolution of the Galactic halo with VST'' (PI: M. Marconi) and by PRIN--MIUR
(2010LY5N2T) ``Chemical and dynamical evolution of the Milky Way and Local Group
galaxies'' (PI: F. Matteucci).  APM acknowledges the financial support from the
Australian Research Council through Discovery Project grant DP120100475.
Support for this work has been provided by the IAC (grant 310394), and the
Education and Science Ministry of Spain (grants AYA2007-3E3506, and
AYA2010-16717).
\end{acknowledgements}



\begin{thebibliography}{}

\bibitem[Anderson \& King(2003)]{And03} Anderson, J., \& King, I.~R.\ 2003, \aj,
126, 772 

\bibitem[Anderson \& King (2006)]{And06} Anderson, J.,\& King, I.~R.\ 2006,
Instrument Science Report ACS 2006-01, 34 pages, 1 

\bibitem[Anderson et al.(2008)]{And08} Anderson, J., et al.\ 2008, \aj, 135,
2055 

\bibitem[Asplund et al.(2009)]{asplund09} Asplund, M., Grevesse, N., Sauval, A.
J., \& Scott, P.\ 2009, \araa, 47, 481 

\bibitem[Baade(1958)]{Baa58} Baade, W. 1958, in Stellar Population, ed. D.J.K.
O'Connell (New York: Interscience), 303  

\bibitem[Bedin et al.(2001)]{Bed01} Bedin, L.~R., Anderson, J., King, I.~R., \&
Piotto, G.\ 2001, \apjl, 560, L75 

\bibitem[Bedin et al.(2005)]{Bed05} Bedin, L.~R., Cassisi, S., Castelli, F.,
Piotto, G., Anderson, J., Salaris, M., Momany, Y., \& Pietrinferni, A.\ 2005,
\mnras, 357, 1038

\bibitem[Bellini et al.(2011)]{Bel11} Bellini, A., Anderson, 
J., \& Bedin, L.~R.\ 2011, \pasp, 123, 622 

\bibitem[Bergbusch \& Stetson(2009)]{Ber09} Bergbusch, P.~A., \& Stetson, P.~B.\
2009, \aj, 138, 1455 

\bibitem[Bertelli et al.(1994)]{Ber94} Bertelli, G., Bressan, A., Chiosi, C.,
Fagotto, F., \& Nasi, E.\ 1994, \aaps, 106, 275 

\bibitem[Boesgaard et al.(2009)]{Boe09} Boesgaard, A.~M., Jensen, E.~E.~C., \&
Deliyannis, C.~P.\ 2009, \aj, 137, 4949 

\bibitem[Bono et al.(2010)]{Bon10} Bono, G., Stetson, P.~B., VandenBerg, D.~A.,
et al.\ 2010, \apjl, 708, L74 

\bibitem[Brasseur et al.(2010)]{Bra10} Brasseur, C.~M., Stetson, P.~B.,
VandenBerg, D.~A., et al.\ 2010, \aj, 140, 1672 

\bibitem[Brogaard et al.(2012)]{Bro12} Brogaard, K., VandenBerg, D.~A., Bruntt,
H., et al.\ 2012, \aap, 543, A106 

\bibitem[Brott \& Hauschildt(2005)]{brott05} Brott, I., \& Hauschildt, P.~H.\
2005, The Three-Dimensional Universe with Gaia, 576, 565

\bibitem[Brown et al.(2005)]{Bro05} Brown, T.~M., Ferguson, H.~C., Smith, E., et
al.\ 2005, \aj, 130, 1693 

\bibitem[Calamida et al.(2009)]{Cal09} Calamida, A., Bono, G., Stetson, P.~B.,
et al.\ 2009, \apj, 706, 1277 

\bibitem[Calamida et al.(2012)]{Cal12} Calamida, A., Monelli, M., Milone, A.~P.,
et al.\ 2012, \aap, 544, A152 

\bibitem[Carretta et al.(2001)]{Car01} Carretta, E., Cohen, J.~G., Gratton,
R.~G., \& Behr, B.~B.\ 2001, \aj, 122, 1469 

\bibitem[Carretta et al.(2009)]{Car09} Carretta, E., Bragaglia, A., Gratton, R.,
D'Orazi, V., \& Lucatello, S.\ 2009, \aap, 508, 695 

\bibitem[Cassisi \& Salaris (1997)]{Cas97} Cassisi, S., \& Salaris, M.\ 1997,
\mnras, 285, 593 

\bibitem[Castellani et. al (1999)]{Cast99} Castellani, V., Degli'Innocenti, S.,
Fiorentini, G., Lissia, M., \& Ricci, B. 1999, Phys. Rep., 281, 309

\bibitem[Castelli \& Kurucz(2003)]{castelli03} Castelli, F., \& Kurucz, R.~L.\
2003, Modelling of Stellar Atmospheres, 210, 20P 

\bibitem[Degl'Innocenti et al.(2008)]{deglinnocenti08} Degl'Innocenti, S., Prada
Moroni, P. G., Marconi, M., \& Ruoppo, A.\ 2008, \apss, 316, 25 

\bibitem[Dell'Omodarme et al.(2012)]{database2012} Dell'Omodarme, M., Valle, G.,
Degl'Innocenti, S., \& Prada Moroni, P. G.\ 2012, \aa, 540, A26

\bibitem[Dotter et al.(2010)]{Dot10} Dotter, A., Sarajedini, A., Anderson, J.,
et al.\ 2010, \apj, 708, 698

\bibitem[Dotter et al.(2011)]{Dot11} Dotter, A., Sarajedini, A., \& Anderson,
J.\ 2011, \apj, 738, 74 

\bibitem[Feltzing \& Johnson (2002)]{Fel02} Feltzing, S., \& Johnson, R.~A.\
2002, \aap, 385, 67

\bibitem[Freeman et al.(2013)]{fre13} Freeman, K., Ness, M., Wylie-de-Boer, E.,
et al.\ 2013, \mnras, 428, 3660 

\bibitem[Gennaro et al.(2010)]{gennaro10} Gennaro, M., Prada Moroni, P. G., \&
Degl'Innocenti, S.\ 2010, \aa, 518, A13 

\bibitem[Gonzalez et al.(2011)]{gon11} Gonzalez, O.~A., Rejkuba, M., Zoccali,
M., et al.\ 2011, \aap, 530, A54 

\bibitem[Hansen et al.(2013)]{han13} Hansen, B.~M.~S., Kalirai, J.~S., Anderson,
J., et al.\ 2013, \nat, 500, 51 

\bibitem[Harris(1996)]{Har96} Harris, W.~E.\ 1996, \aj, 112, 1487

\bibitem[Jimenez et al.(2003)]{jimenez03} Jimenez, R., Flynn, C., MacDonald, J.,
\& Gibson, B. K.\ 2003, \science, 299, 1552 

\bibitem[Kuijken \& Rich(2002)]{Kui02} Kuijken, K., \& Rich, R.~M.\ 2002, \aj,
124, 2054
 
\bibitem[Mar{\'{\i}}n-Franch et al.(2009)]{Mar09} Mar{\'{\i}}n-Franch, A.,
Aparicio, A., Piotto, G., et al.\ 2009, \apj, 694, 1498

\bibitem[McLaughlin et al.(2006)]{McL06} McLaughlin, D.~E., Anderson, J.,
Meylan, G., et al.\ 2006, \apjs, 166, 249

\bibitem[McWilliam \& Rich(1994)]{mcw94} McWilliam, A., \& Rich, R.~M.\ 1994,
\apjs, 91, 749 

\bibitem[McWilliam \& Zoccali(2010)]{McW10} McWilliam, A., \& Zoccali, M.\ 2010,
\apj, 724, 1491

\bibitem[Milone et al.(2009)]{Mil09} Milone, A.~P., Bedin, L.~R., Piotto, G., \&
Anderson, J.\ 2009, \aap, 497, 755

\bibitem[Milone et al.(2012)]{Mil12} Milone, A.~P., Piotto, G., Bedin, L.~R., et
al.\ 2012, \aap, 540, A16

\bibitem[Momany et al.(2003)]{Mom03} Momany, Y., Ortolani, S., Held, E.~V., et
al.\ 2003, \aap, 402, 607 

\bibitem[Ness et al.(2013a)]{nes13a} Ness, M., Freeman, K., Athanassoula, E., et
al.\ 2013, \mnras, 430, 836 

\bibitem[Ness et al.(2013b)]{nes13b} Ness, M., Freeman, K., Athanassoula, E., et
al.\ 2013, \mnras, 432, 2092 

\bibitem[Origlia et al.(2005)]{Ori05} Origlia, L., Valenti, E., \& Rich, R.~M.\
2005, \mnras, 356, 1276 

\bibitem[Ortolani et al.(1995)]{Ort95} Ortolani, S., Renzini, A., Gilmozzi, R.,
et al.\ 1995, \nat, 377, 701 

\bibitem[Ortolani et al.(2001)]{Ort01} Ortolani, S., Barbuy, B., Bica, E., et
al.\ 2001, \aap, 376, 878 

\bibitem[Pagel \& Portinari (1998)]{pagel98} Pagel, B.  E. J.\& Portinari, L.\
1998, \mnras, 298, 747 

\bibitem[Peimbert et al.(2007)]{peimbert07} Peimbert, M., Luridiana, V., \&
Peimbert, A.\ 2007, \apj, 666, 636 

\bibitem[Rakos \& Schombert (2005)]{Rak05} Rakos, K., \& Schombert, J.\ 2005,
\pasp, 117, 245 

\bibitem[Salaris et al.(2007)]{Sal07} Salaris, M., Held, E.~V., Ortolani, S.,
Gullieuszik, M., \& Momany, Y.\ 2007, \aap, 476, 243 

\bibitem[Salasnich et al.(2000)]{Sal00} Salasnich, B., Girardi, L., Weiss, A.,
\& Chiosi, C.\ 2000, \aap, 361, 1023 

\bibitem[Steigman(2006)]{steigman06} Steigman, G.\ 2006, \ijmpe, 15, 1

\bibitem[VandenBerg et al.(2010)]{Van10} VandenBerg, D.~A., Casagrande, L., \&
Stetson, P.~B.\ 2010, \aj, 140, 1020 

\bibitem[VandenBerg et al.(2013)]{Van13} VandenBerg, D.~A., Brogaard, K.,
Leaman, R., \& Casagrande, L.\ 2013, \apj, 775, 134 

\bibitem[van Tulder(1942)]{vant} van Tulder, J. J. M. 1942, BAN, 9, 315

\bibitem[Zoccali et al.(2001)]{Zoc01} Zoccali, M., Renzini, A., Ortolani, S.,
Bica, E., \& Barbuy, B.\ 2001, \aj, 121, 2638

\bibitem[Zoccali et al.(2004)]{Zoc04} Zoccali, M., Barbuy, B., Hill, V., et al.\
2004, \aap, 423, 507 

\bibitem[Zoccali et al.(2006)]{Zoc06} Zoccali, M., Lecureur, A., Barbuy, B., et
al.\ 2006, \aap, 457, L1 

\end{thebibliography}
\end{document}